\documentclass[a4paper]{spie}  
\usepackage[]{graphicx}

\title{Layer Oriented wavefront sensor for MAD
on sky operations}

\author{C. Arcidiacono\supit{a}, M. Lombini\supit{b}, R. Ragazzoni\supit{a}, J. Farinato\supit{a}, E. Diolaiti\supit{b},  A. Baruffolo\supit{a}, P. Bagnara\supit{a}, G. Gentile\supit{a}, L. Schreiber\supit{b}, E. Marchetti\supit{c}, J. Kolb\supit{c}, S. Tordo\supit{c}, R. Donaldson\supit{c}, C. Soenke\supit{c}, S. Oberti\supit{c}, E. Fedrigo\supit{c},E. Vernet\supit{c}, N. Hubin\supit{c}
\skiplinehalf
\supit{a}INAF Osservatorio Astronomico di Padova, Vicolo dell'Osservatorio 5 I-35122, Padova, Italy; \\
\supit{b}INAF Osservatorio Astronomico di Bologna, Via Ranzani 1 I-40127, Bologna, Italy; \\
\supit{c}European Southern Observatory, Karl Schwartzchild strasse 2, D-85748 Garching bei  M\"unchen, Germany
}

\authorinfo{Further author information: (Send correspondence to Carmelo Arcidiacono)\\E-mail: carmelo.arcidiacono@oapd.inaf.it, Telephone: +39 049 829 3523\\              }

\begin{document}
  \def\arcmin{$^{\prime}$}
  \def\arcsec{$^{\prime\prime}$}
  \def\arcdeg{$^{\rm o}$}
  \def\Ib{{\bf I} (0.85 $\mu${\em m})$\;$}
  \def\Jb{{\bf J} (1.25 $\mu${\em m})$\;$}
  \def\Hb{{\bf H} (1.65 $\mu${\em m})$\;$}
  \def\Kb{{\bf K'} (2.12 $\mu${\em m})$\;$}
  \def\byone{1\arcmin$\times$1\arcmin}
  \def\bytwo{2\arcmin$\times$2\arcmin}
  \maketitle

\begin{abstract}
The Multiconjugate Adaptive optics Demonstrator (MAD) has successfully demonstrated on sky both Star Oriented (SO) and Layer Oriented (LO) multiconjugate adaptive optics techniques. While SO has been realized using 3 Shack-Hartmann wavefront sensors (WFS), we designed a multi-pyramid WFS for the LO. The MAD bench accommodates both WFSs and a selecting mirror allows choosing which sensor to use. In the LO approach up to 8 pyramids can be placed on as many reference stars and their light is co-added optically on two different CCDs conjugated at ground and to an high layer. In this paper we discuss LO commissioning phase and on sky operations.
\end{abstract}


\keywords{Adaptive Optics, Multiconjugate Adaptive Optics, High Angular Resolution}

\section{The Multiconjugate Adaptive Optics Demonstrator}
\label{sec:intro}  
Adaptive optics (AO) is a technique which provides real time correction of the optical aberration generated by atmospheric turbulence.
The Multiconjugate Adaptive optics Demonstrator\cite{hubin02,MADFDR} (MAD) is the ESO experiments which successfully demonstrated the feasibility of the Multiconjugate Adaptive Optics\cite{beckers88,beckers89a} technique (MCAO) on sky. MCAO aims to overcome the anisoplanatism problem of the single conjugate adaptive optics correction (SCAO), currently performed on different observatories, improving in this way also the angular dimension of the corrected field and, at the end, the sky coverage. In fact SCAO provides correction only on a limited angular dimension (10arcsec-20arcsec) because the atmospheric volume seen by the wavefront sensing system is limited to the direction of the reference guide star. MCAO senses simultaneously the wavefronts of several guide stars reconstructing the three dimensional distribution of the atmospheric optical aberration. On a MCAO instrument at least two deformable mirrors (DM) are optically conjugated to as many atmospheric layers realizing in this way the isoplanatic correction. But it is possible to use a single deformable mirror to perform the correction of the ground, and most turbulent, layer realizing in this way the so called Ground Layer Adaptive Optics (GLAO), which provides less efficient correction but can be used on much larger FoV (2\arcmin to 6\arcmin) using several guide stars.

On MAD two different MCAO approaches have been implemented: the Star Oriented (SO) multi Shack Hartmann\cite{marchetti06} wavefront sensor (WFS) and Layer Oriented\cite{LO1,LO2} (LO) multi pyramid {WFS\cite{ragazzoni02,viard2003,arcidiacono07}}. Both sensors look for reference stars on a 2\arcmin field of view: the slopes measurements derived by the WFS are then translated by a wavefront computer to voltages for the two deformable mirrors optically conjugated to 0 and 8.5 km from telescope pupil. The CAMCAO\cite{CAMCAO1} scientific infrared camera has been built by the Universidad de Lisboa, it is mounted on the corrected focal plane of MAD: CAMCAO is a high resolution, wide Field of view NIR camera, that uses the 2k$\times$2k HgCdTe HAWAII-2 infrared detector from Rockwell Scientific, controlled by the ESO IRACE system. The camera operates in
the near infrared region between 1.0$\mu$m and 2.5 $\mu$m wavelength using a filter wheel with J, H,
K', K-continuum and Br$\gamma$ filters.
\begin{figure}
\centerline{\includegraphics[height=2.8in]{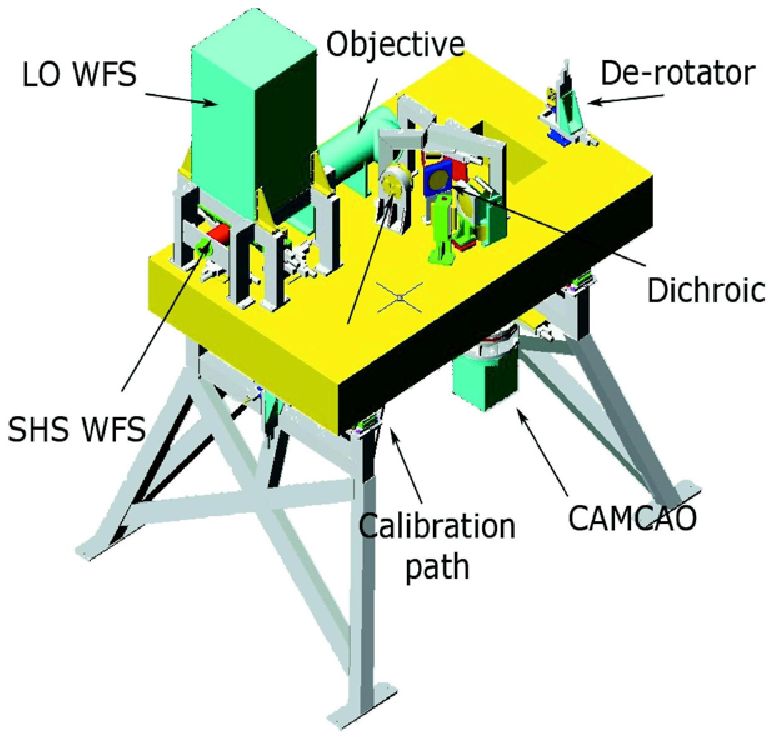}\includegraphics[height=2.8in]{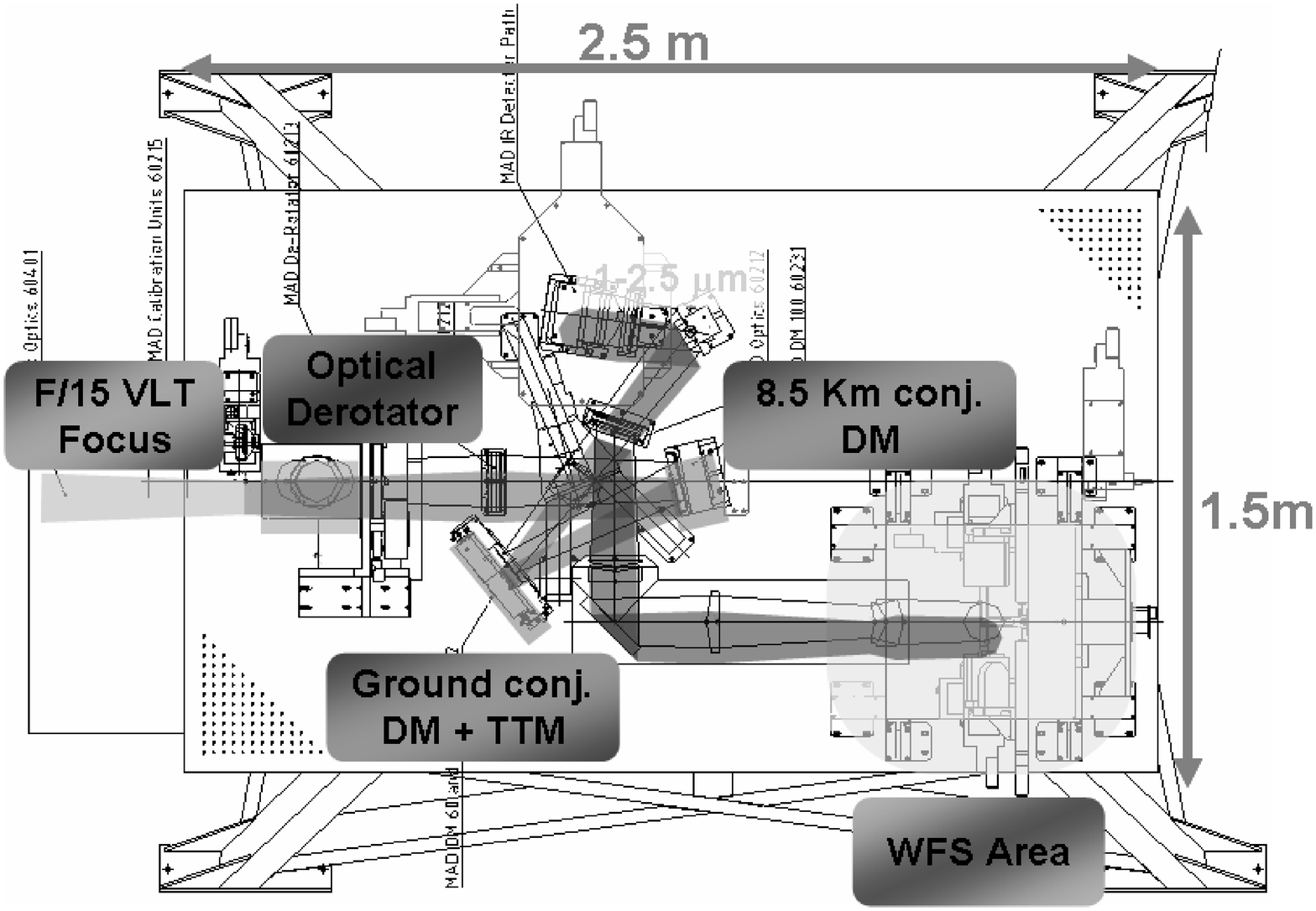}}\caption{\footnotesize
The CAD view on the left presents the MAD bench and the instrument main 
components of the mechanical design concept: the beam arrives on
the de-rotator, passes through the collimator and is successively
reflected by the two Deformable Mirrors (DMs). Then the IR light is transmitted to the
camera while the visible light is reflected by the dichroic toward
the WFS objectives. The Multi Shack-Hartmann WFS  is located below the LO--WFS (the elongated box
on the left side). On the right a top view.}\label{fig:madbench}
\end{figure}
The MAD--bench is not fixed to the VLT--Nasmyth adapter rotator flange, then the pupil rotates with the field: an optical de--rotator at the entrance of the adaptive--system rotates both.
The adaptive system is illuminated by a re--imaging optic collimating the F/15 input beam in order to re--image the telescope pupil on the ground layer bimorph deformable mirrors and conjugating the second to 8.5 km (both DM are XINETICS with 60 actuators, but high DM is slightly larger, 100 mm diameter). The infrared (IR)
light is transmitted by the dichroic to the CAMCAO scientific IR camera while the
visible light is reflected toward the WFS path. The MAD bench common optics retrieves to the WFS a flat telecentric F/20 input beam.

MAD has been mounted on the Visitor Focus on one of the Nasmyth platforms of the VLT-Melipal (UT3) in 2007 and between 21$^{\rm st}$ to 29$^{\rm st}$ September the telescope has been scheduled for 3 Technical and 6 Guaranteed Time Observations (GTO) nights in Layer Oriented mode.
In this paper we will discuss the most important technical results obtained during these nights.

\section{The Layer Oriented Wavefront Sensor}
\begin{figure}[htbp]
\centerline{
\includegraphics[height=3.0in]{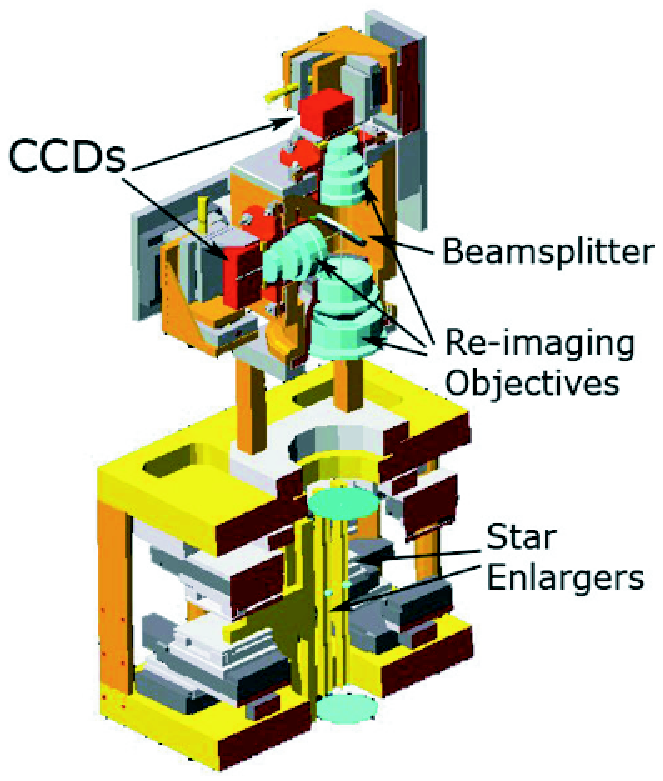}\label{fig:madcad}
\includegraphics[height=3.0in]{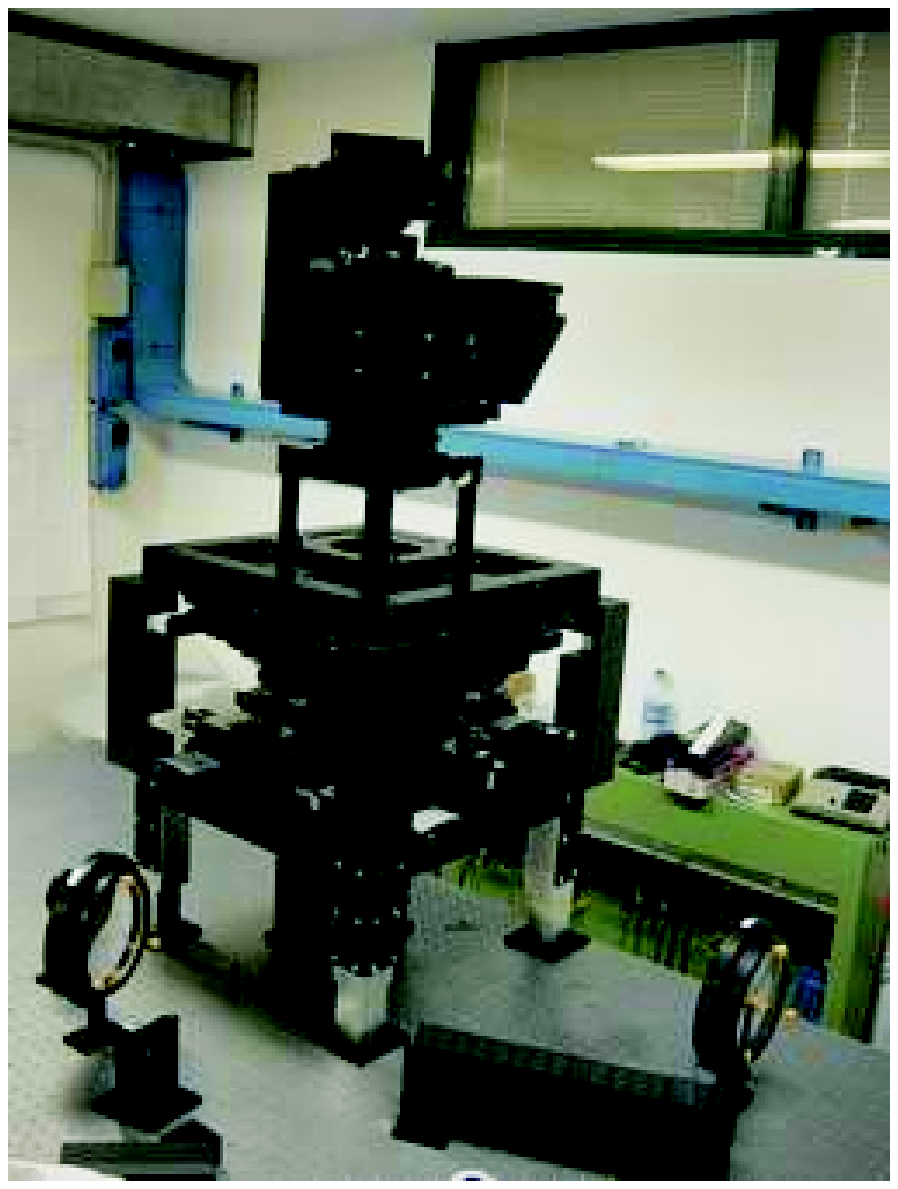}\label{fig:madphoto}}
\caption{\footnotesize These figures present the CAD
view and a real picture
of the MAD Layer Oriented WFS.}
\label{fig:mad3d}
\end{figure}
\begin{figure}
\centerline{\includegraphics[width=18cm]{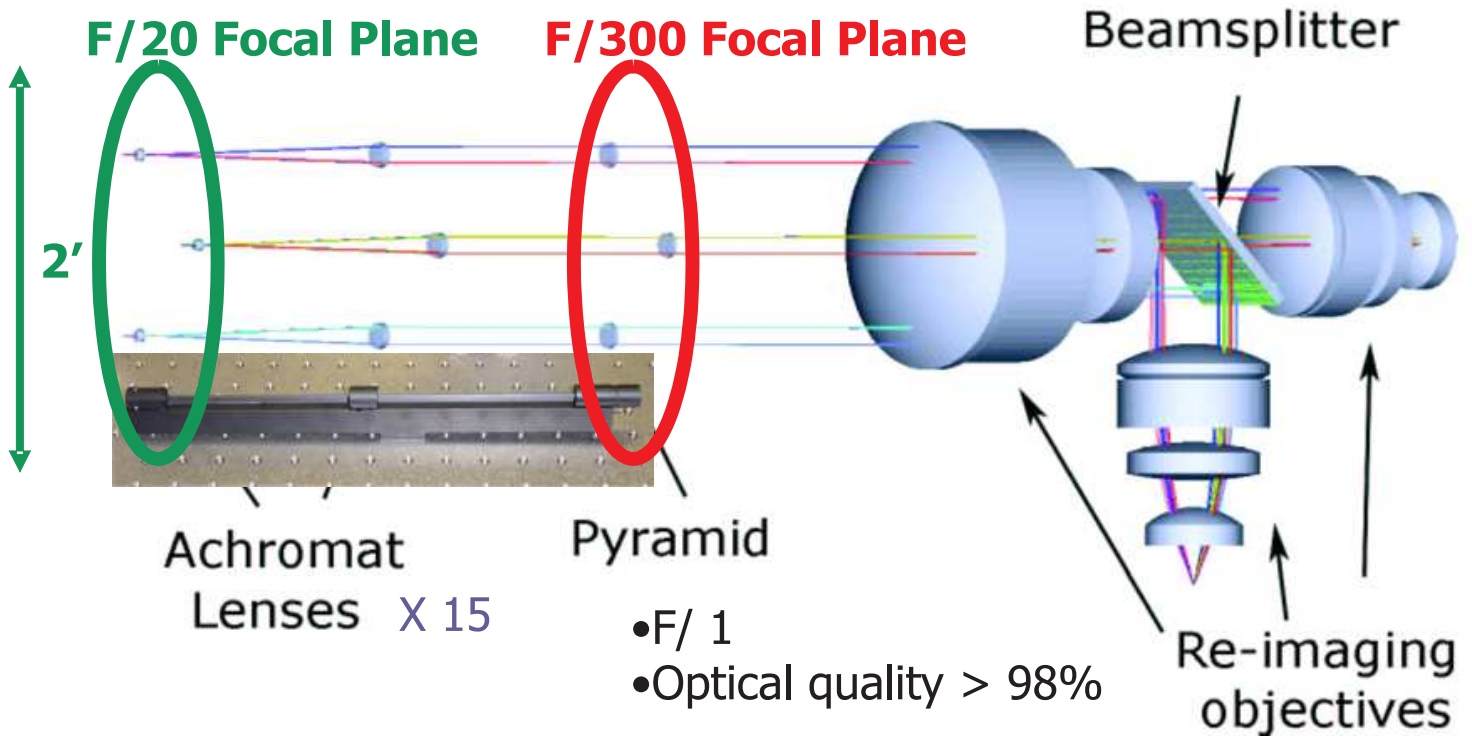}}\caption{\footnotesize
The optical layout of the LO--WFS. In the real system 8 pyramids
are available even if only 3 are shown here. The stars light coming from the left is caught by
the system composed by the two achromatic doublets and split by the pyramids. The beams pass through the common path pupil
re--imager before to be split to the high and the ground pupil
re--imaging objectives. The pupil images are sensed by the two
CCDs.}\label{fig:madopt3d}
\end{figure}

In the LO--WFS for MAD the multi--pyramids approach is exploited: as in single reference {case\cite{pyramid,pyrmodal}}, the light of each reference stars is split into four beams by a pyramid placed on a focal plane and centered on the star. Then a reimaging objective projects on a CCD the pupil images (4 pupil images, one for each pyramid face), which are super--imposed according to the conjugation altitude and the stars directions.
In the LO--WFS for MAD up to 8 pyramids can be positioned over the
2\arcmin FoV to catch the light from eight reference guide
stars, Fig.~\ref{fig:mad3d}. In the Layer Oriented\cite{LO1,LO2} approach the pupil images are co--added mimicking the super--imposition of the reference stars beams through the atmosphere as they do at the conjugation altitude. In MAD the reference pupils co--addition is optically performed on two different WFSs, which looks simultaneously to the four pupil images of all reference stars. A beam splitter placed on the pupil re--imager optical path splits the light to two identical objectives. The pupil image corresponding to the F/20 WFS--input beam is too large to fit the dimension of the wavefront sensing fast read out CCD, even if used with very fast re--imaging optics.
In fact the detector size is one crucial
parameter to determine the optics characteristics and the final LOWFS
optical design was decided to fit this size. The two
detectors are EEV39 with $80\times 80$ pixels, with 0.024mm pixel--size corresponding to a $1.92\times 1.92$mm sensing region.
Pupil size shrinking was necessary to fit CCD size: in general this can be achieved by enlarging the focal ratio before the pupil re--imager, but in this way the focal plane corresponding to the 2\arcmin field is enlarged of the same factor and the re--imager aperture too. The trick used to still use small re--imaging optics is described in the cited paper\cite{tubetti} and it consists in enlarging the focal ratio only on the optical path of the reference stars in front of each pyramid. This goal was achieved using a co--moving optical train for each pyramid: two small diameter achromatic doublets have been used to obtain a new F/300 focal plane in correspondence of the
pyramids vertex, Fig.~\ref{fig:madopt3d}. The second lens diameter fixes the minimum
separation between two reference stars: because of their
physical dimensions two adjacent pyramids cannot be more
close then a distance corresponding to 20\arcsec (centre--centre)
if projected on sky. In this way the pupil has been shrink by a factor 15, and using a very fast F/1.05 re--imager, an $\sim 0.388 mm$ pupil size has been obtained. Moreover as small as possible divergence angle for the pyramids have been selected ($\approx 1.2deg$ opposite pyramid faces tilt angle \cite{arcidiacono}) to accommodate the four pupils on the CCD.
Using a fixed $2\times 2$ binning for ground CCD and $4\times 4$ for high one the metapupil spatial sampling obtained was respectively $8\times 8$ and $7\times 7$.

The two CCDs are positioned with great accuracy ($\approx 1 \mu m$) in order to properly conjugate to deformable mirrors and corresponding layer altitudes. The CCD are mounted on linear stages movements to correct the position of the conjugated planes. The motorized stages are the same used to move over the 2\arcmin FoV the mounting of the optical train (called Star Enlarger) composed by the pyramid and the 2 achromatic doublets: in fact these 8 opto-mechanical systems are screwed to as many xy couples of linear stages moving on a plane orthogonal to the optical axis.

\section{Sky Operations}
\label{sec:ope}
During both technical and GTO nights we had a list of possible targets to observe.
The technical nights have been devoted to evaluate the correction performance of the Layer Oriented wavefront sensor under different system configurations and to bring the Layer Oriented system to a level of functionality sufficient to properly exploit the GTO nights (from September 24$^{\rm th})$. Thus the observations during this time have been purely technical. Based on the experience of the previous Star Oriented commissioning\cite{bouy08,bono08} the selected targets consisted mainly in crowded fields including several bright guide stars up to V$\approx$10.

\subsection{The reference stars acquisition procedure}
\label{sec:acq}
In order to proceed to a new target observation several operations were necessary. In fact we already known from laboratory tests that the positioning of the pyramids on the reference stars over the field is a critical point to obtain best performance. On MAD is possible to have a pre-view of the observed 2\arcmin FoV thanks to an optical camera (called Technical CCD) which allows to identify and select the reference stars through a direct imaging of the whole two arcmin FoV. But Technical CCD is behind the folding mirrors which has to be positioned on the optical axis in order to reflect the light to one of the two WFS or to be removed to illuminate this camera. A template has been developed to speed up the operation that anyway takes a long time because of the long travel range ($\sim$ 40mm) of the two folding mirrors linear stages ($\sim$ 1mm/sec). Then pyramids are positioned one by one on the reference stars: their centering is adaptively adjusted by correcting the liner stages positions by minimizing the tip and tilt signals measured on the WFS. Such as procedure posed a limitation on the brightness of the reference stars being the signal corresponding to sky and RON equivalent to a 14.12 $\pm0.2$ magnitude star. The slowest CCD readout mode implemented on MAD was the 50Hz: by reaching slower frequency in principle could be possible to use fainter stars then the ones used. For the faintest stars used we had to average frames for about 30-40seconds in order to reach a sufficient signal to noise. The effect of slow linear stages and fast readout modes translated on acquisition time of the order of the 15-20minutes.
\subsection{Closing AO loop}
\label{sec:close}
While reference stars acquisition procedure took long time the procedure to close the loops was quite fast, needing only to load the interaction matrix recorded in daytime using fiber sources and to select the number of modes to be used by setting to zero the eigenvalues corresponding to the unused modes for matrix inversion. In fact one of the advantage of LO is that Interaction Matrix to be inverted to generate Control Matrix is not dependent on the particular reference stars constellation used. Once loaded the Control Matrix the loops can be closed and user can proceed to the fine tuning of the gains applied respectively to ground and high loop. Should be noticed that actually we used both loops independently (Sensor``Ground" driving only ground DM and sensor ``High" driving only the 8.5km one) and also with a mixed control matrix (using also non diagonal terms on Interaction and Control Matrix). In laboratory it was possible to promote the mixed control matrix such as most performing but was not possible to confirm this result on sky because of the poor statistic, poor seeing and its large temporal variations.
In MCAO closed loop the removal from the slope measurements vector of the low flux sub-apertures was fundamental to obtain best performance.

\section{Single Conjugate Adaptive Optics}
The 21$^{\rm st}$ we started the operations with a single reference star in order to check the functionality of the system. We used a V$\approx 9.47$ magnitude star ($\alpha = $22 01 09.10, $\delta = $-41 18 23.363). The system was working properly and so we proceeded to optimize the loop gain and the number of modes. Moreover we tested   different Control Matrices obtained inverting Interaction Matrices obtained with and without tip-tilt modulation, registered using all pyramids on as many reference fibers or using only the best aligned ones. As results we obtained that the most performing number of modes is 40 over 60 and that whenever possible is better to use interaction matrix recorded using the same pyramid used for the the observation in order to reduce the effect of the mis-alignment introduced by the wobbling of the linear stage moving the pyramid optical train over the field. Best Strehl Ratio obtained was 34.8\% with a seeing measured on the camera on the same $Br\gamma$
filter of 0.35\arcsec.
Since system was working as expected we pass to a fainter SCAO case using a V$\approx$12.47 magnitude star, as measured by the WFS ($\alpha$=330.21774, $\delta$=-41.29427), USNO  B1 Catalogue gives for this star B=12.8 and R=11.9. We observed using the best of the previously tested pyramids and only using non modulated recorded Interaction Matrix.
\begin{figure}
\centerline{\includegraphics[width=10cm]{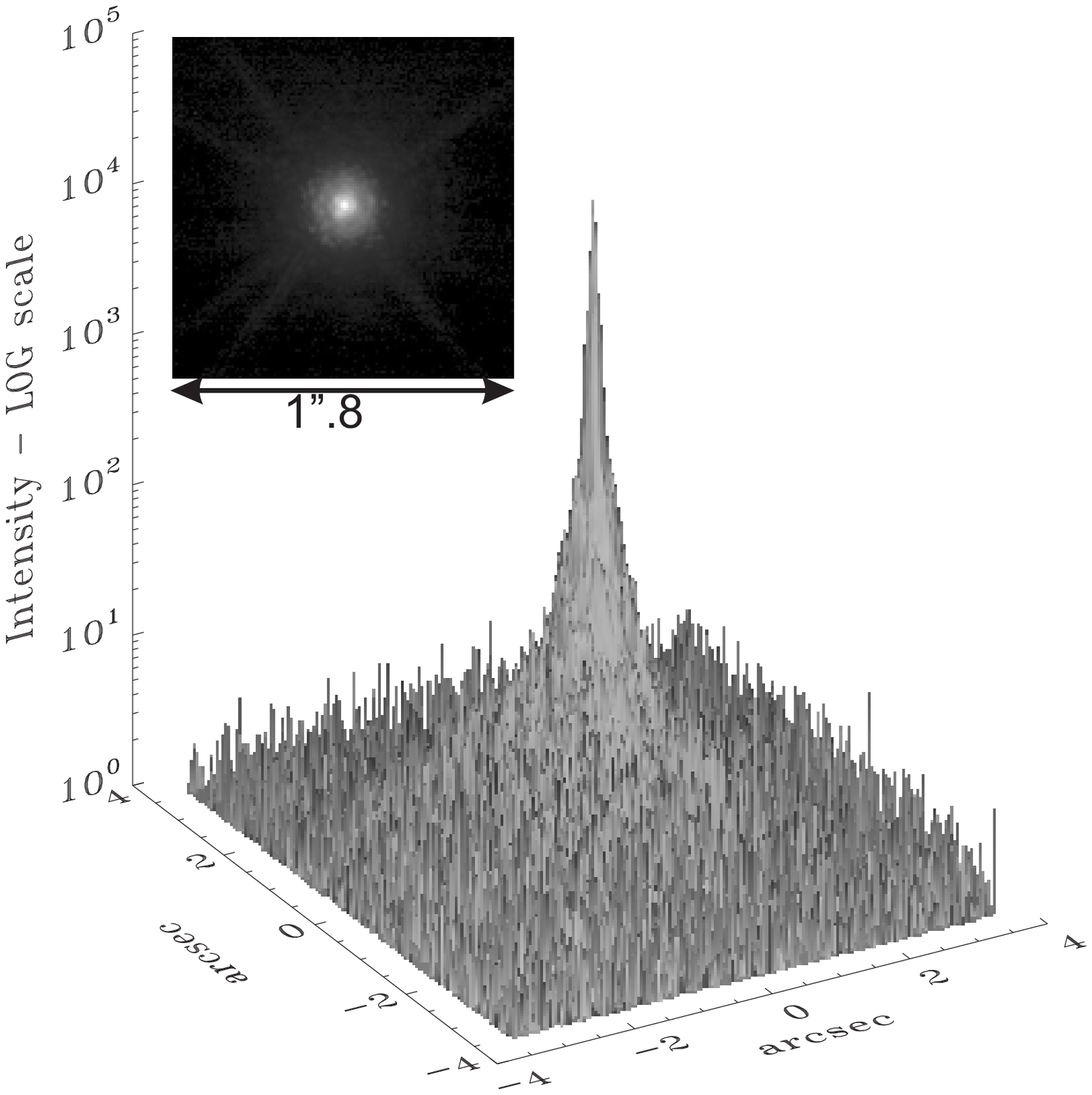}
\includegraphics[width=10cm]{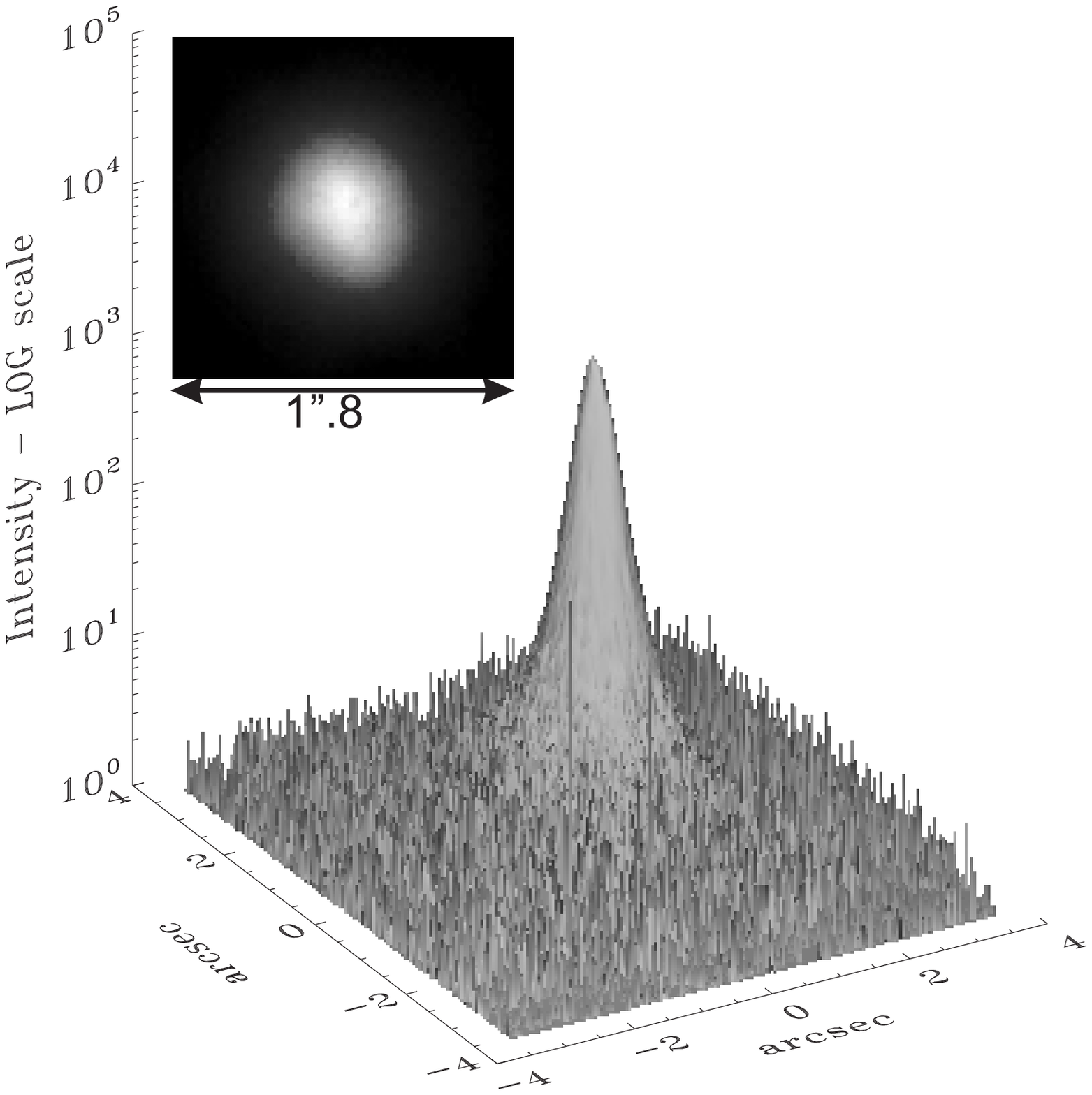}}\caption{\footnotesize
The surface shows the PSF relative to the best SR performed (34.8\%), measured in the $Br\gamma$ filter, obtained using 40 modes and 20\% loop gain. A pyramid has been positioned on a V$\approx$9.47 star and loop has been closed applying an interaction matrix recorded in daytime using this particular star enlarger alone. In open loop the FWHM measured on the PSF imaged by the CAMCAO camera was 0.35\arcsec.}\label{fig:scao}
\end{figure}

\begin{table}[h]
\caption{The following table summarizes the results obtained in SCAO mode. Full width at half Maximum, FWHM, are measured by fitting Moffat Function.  The Ensquared Energy in 0.1\arcsec (EE$_{0.1}$) and 0.2\arcsec (EE$_{0.2}$) are listed showing a gain $\sim \times 4$ and $\sim \times 3$ with respect to open loop case. SR$_{{\rm Br}}\gamma$ and $\sigma_{{\rm{V,DIMM}}}$ are respectively the measured Strehl Ratio in ${{\rm Br}}\gamma$ and the seeing FWHM measured by the Paranal DIMM seeing monitor during the exposures and measured in the V-band. The images in closed loop are exposures of 30sec, while the open loop one is 60sec.}
\label{tab:ScaoB}
\begin{center}
\begin{tabular}{|l|r|r|r|r|r|r|r|} 
\hline
\rule[-1ex]{0pt}{3.5ex}              & M$_{V}$         & FWHM [\arcsec] ${{\rm Br}}\gamma$ & EE$_{0.1}$ [\%] ${{\rm Br}}\gamma$& EE$_{0.2}$ [\%] ${{\rm Br}}\gamma$& SR$_{{\rm Br}}\gamma$ [\%]& $\sigma_{{\rm{V,DIMM}}}$ [\arcsec] @V\\
\hline
\rule[-1ex]{0pt}{3.5ex}  Bright Closed Loop &  9.47    & 0.07 & 33.6       &  60.2      & 34.8          &  0.921\arcsec            \\
\hline
\rule[-1ex]{0pt}{3.5ex}  Faint Closed Loop  &  12.47   & 0.08 & 24.3       &  50.0      & 26.3          &  0.853\arcsec            \\
\hline
\rule[-1ex]{0pt}{3.5ex}  Open Loop          &  -       & 0.35 &        7.7 & 20.5       & 1.7           &  0.886\arcsec            \\
\hline
\end{tabular}
\end{center}
\end{table}

\section{Ground Layer Adaptive Optics}
The following step was the realization of the Ground Layer Adaptive Optics correction. In order to avoid problems related to photon noise on the WFS we selected a bright reference stars constellation in the globular cluster 47 Tuc (NGC 104). Nevertheless the large airmass ($\sim 1.7$) it was a good target thanks to not so bad seeing conditions (around 1.2\arcsec, by DIMM in the V filter) and the bright constellation.
We acquired a crowded field centered on the globular cluster pointing the telescope at $\alpha$ = 6.023634, $\delta$= -72.080410. The goal was to optimize the GLAO loop and have a first estimation of the performance using one mirror only before to go to MCAO. In fact because of Layer Oriented approach the GLAO control matrix is a portion of the MCAO one, and the optimization of this configuration is half way to configure MCAO case. After gain and modes optimization the seeing measured by Dimm seeing monitor was slightly worse $\sim$1.5arcsec (in V band) and we measured 0.45arcsec FWHM in open loop in Bracket-$\gamma$ (2.166$\mu m$).
We used 4 reference stars, well spaced over the field: magnitudes were 11.9, 11.9, 12.4 and 12.5 corresponding to an integrated magnitude 10.63, as they have been measured using the tool available on the instruments itself (respectively for pyramid unit 5, 7, 1 and 3). Using the same tool and the same stars on the night of 23$^{\rm rd}$ Sept. the resulting integrated magnitude was 10.7. But the magnitude has been estimated also using the real time download panel recording flux data: the average flux recorded at 50Hz gives 11.06 magnitudes (which became 10.77 correcting for central obstruction illumination $\epsilon = 0.37$) using 27.398 as Zero Point (ZP). A second set of measurements has been taken giving 10.76 (without using $\epsilon$correction). The difference is due to the different seeing value: in fact a large seeing moves starlight outside the star enlargers-pyramids Field of View, which is only 0.92\arcsec.
The difference 11.06-10.7=0.34 is the differential value in term of ZP of the control panel tool and our flux total magnitudes ZP estimation.

During the observations we noticed that an adjustment of the pyramids positions on the guide reference stars was needed. Later we discovered that a small error in the derotator generated a rotation of the field with respect to optical axis and then with respect to the pyramids vertex. Unfortunately rotation cannot be corrected by the GLAO and MCAO correction (while can easily corrects for tracking errors). Some under-performance was due to this mis-positioning of the pyramids with respect to the reference stars light barycenters especially when close to the Zenith where the derotator speed is larger.

\begin{figure}
\centerline{\includegraphics[width=8cm]{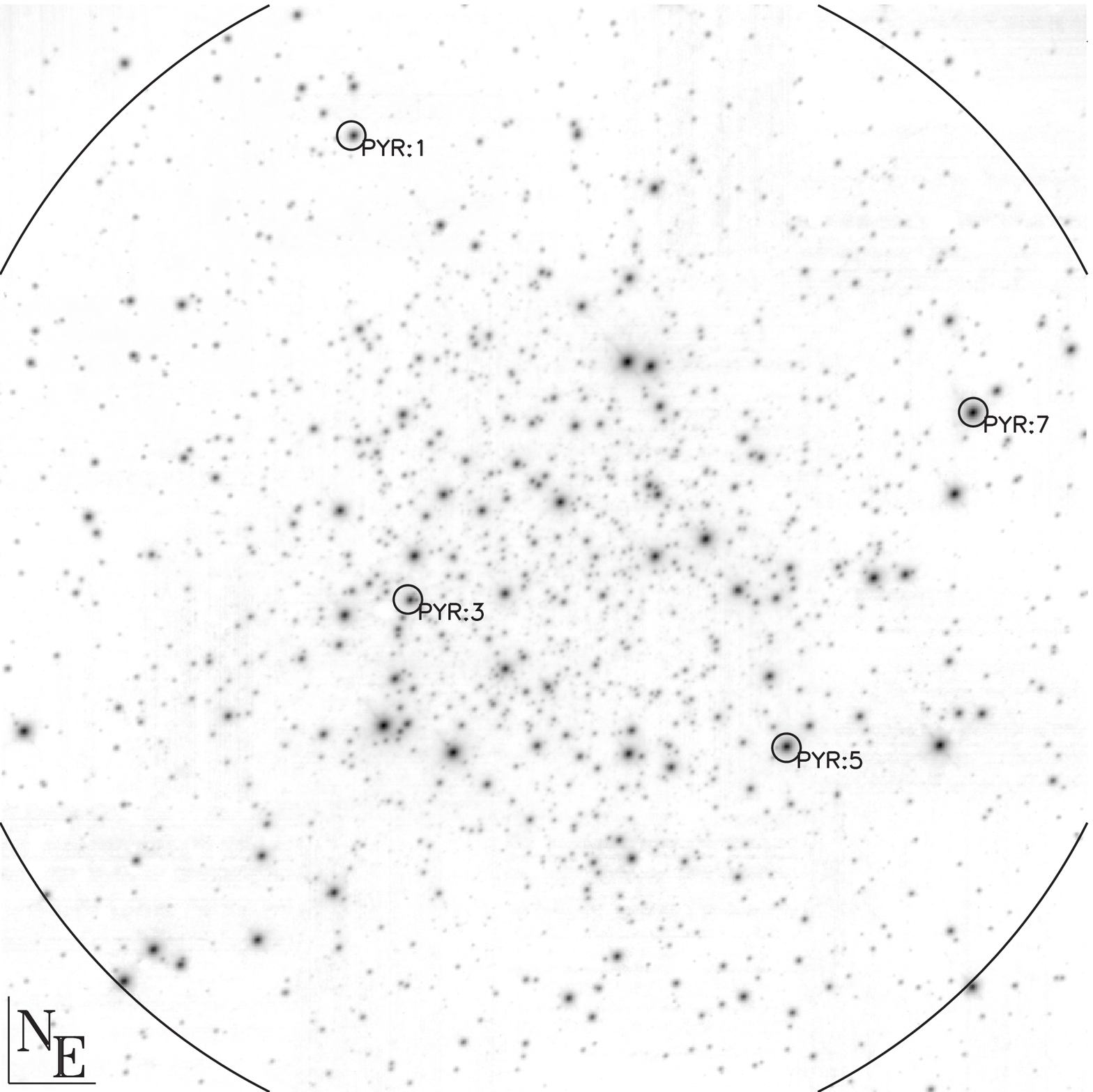}
\includegraphics[width=8cm]{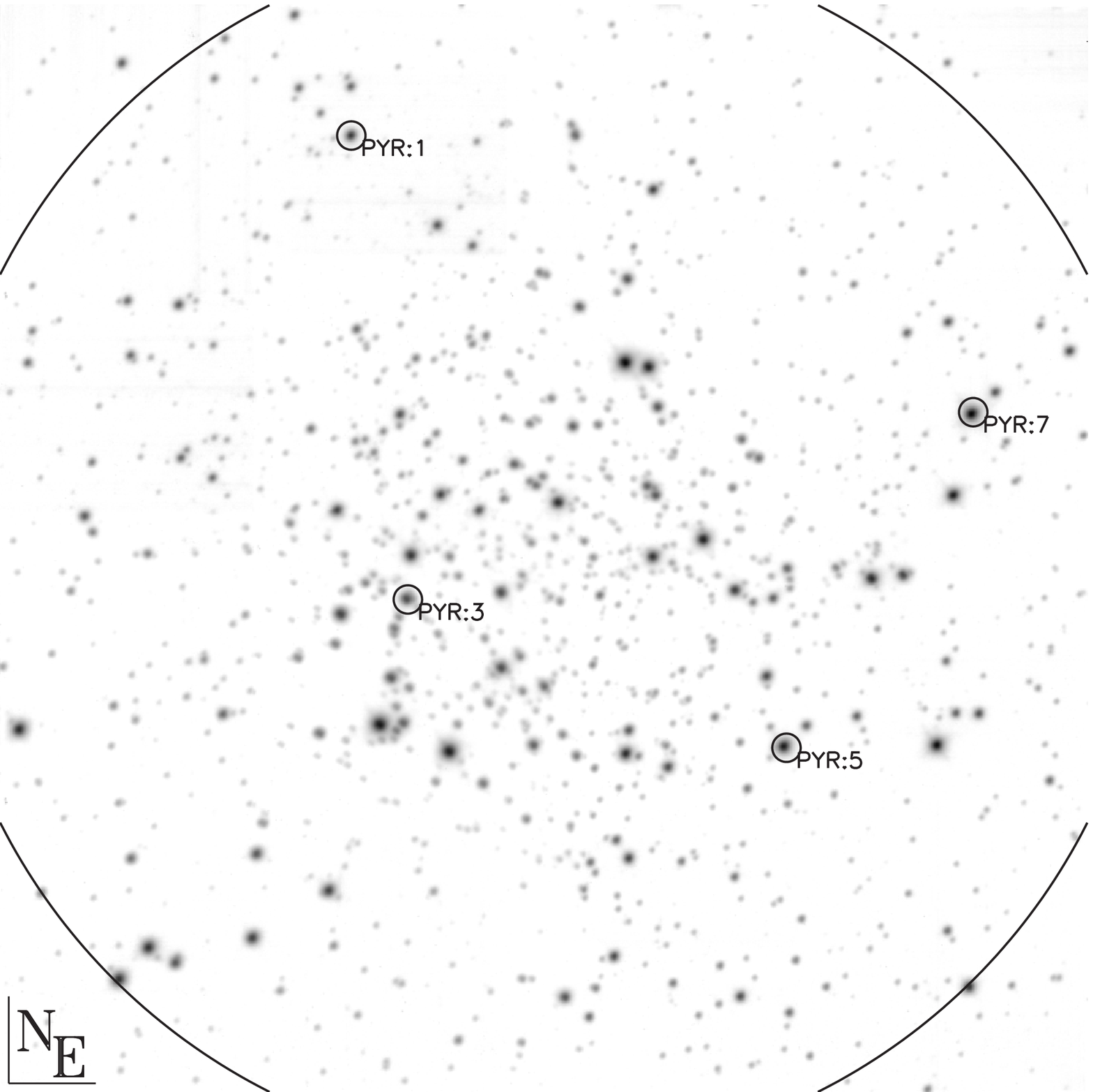}}\caption{\footnotesize
The picture on the left is 107\arcsec$\times$107\arcsec mosaic observed in GLAO closed loop while on the right the same field but in open loop. The circle big represents the corrected 2 arcmin FoV. The frame is a composition of 5 frames one in the center and 4 on a 50\arcsec square, 30 sec exposure each. The small circles indicate the 4 reference stars used.}\label{fig:glao}
\end{figure}

\begin{figure}
\centerline{\includegraphics[width=8cm]{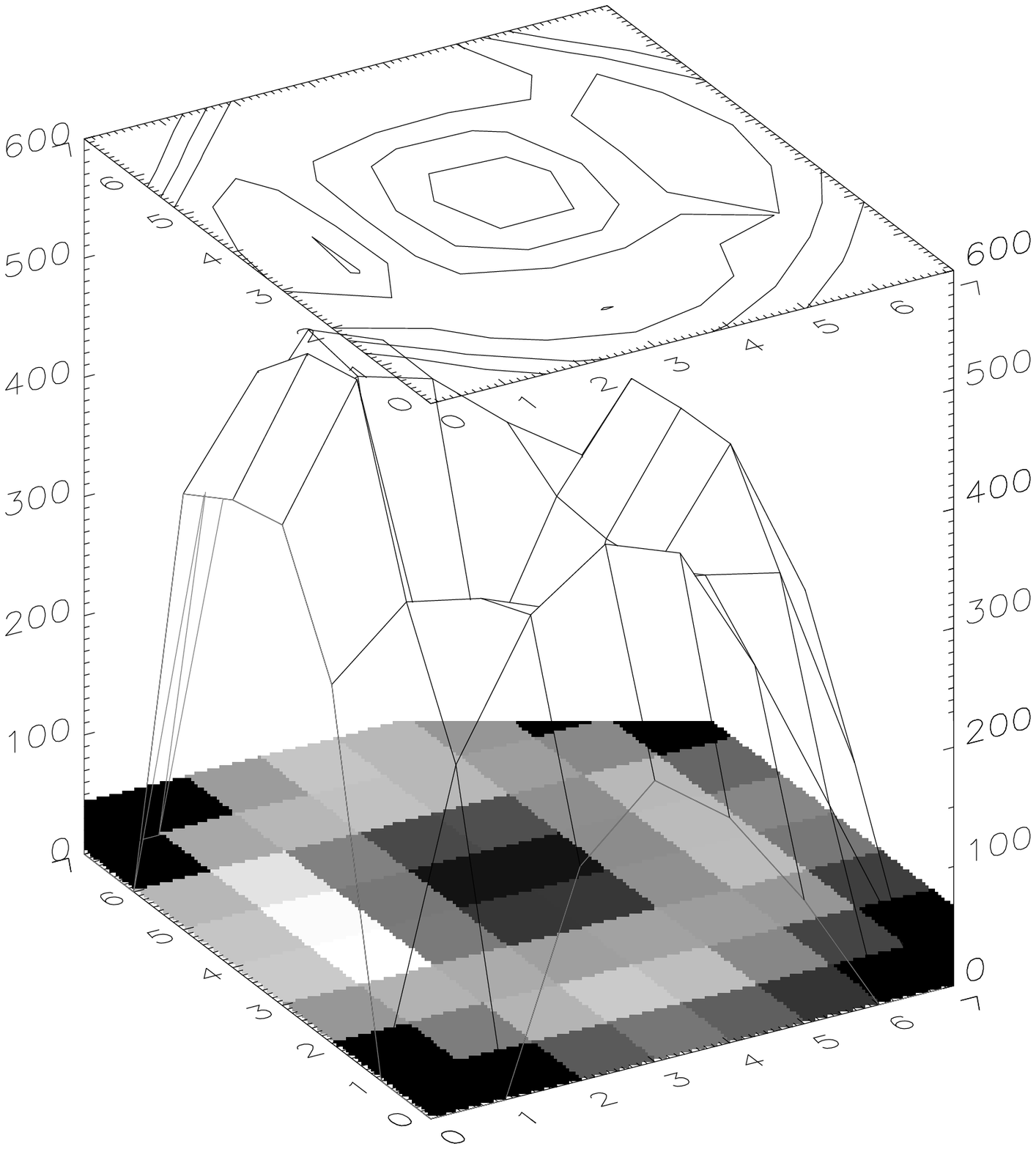}
\includegraphics[width=8cm]{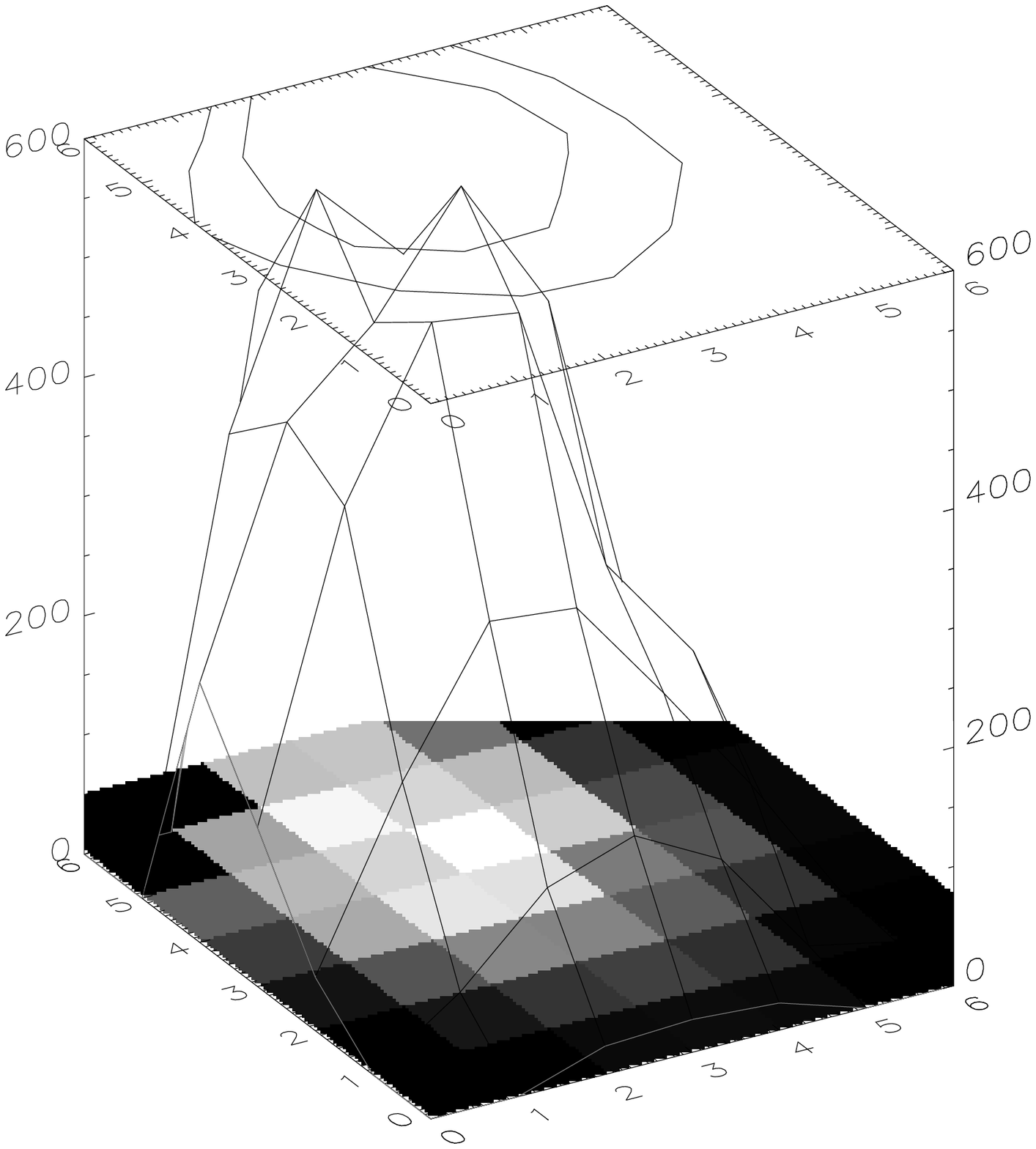}}\caption{\footnotesize
Ground and High WFS pupil illumination for the 47Tuc GLAO case. The surface represents the average counts over the four metapupils. Each data dot shows the counts sum over the 4-pupils for each sub-aperture. The whole illumination is corresponding to an integrated magnitude 11.06 (using as zero point 27.398. This value agrees to the 10.6 coming from the sum of the magnitudes of each reference stars separately measured (11.9, 11.9, 12.4 and 12.5). Flux read at 200 Hz.}\label{fig:subapill}
\end{figure}

We optimized the SR value observing in a \byone region between the four reference stars. But because of large seeing variations was not possible to realize the final \bytwo mosaic at the top of the performance. In the Table~\ref{tab:GLAO0} are presented both open loop and closed loop \bytwo mosaic and the best performance achieved (but measured only on \byone).

\begin{figure}
\centerline{\includegraphics[width=8cm]{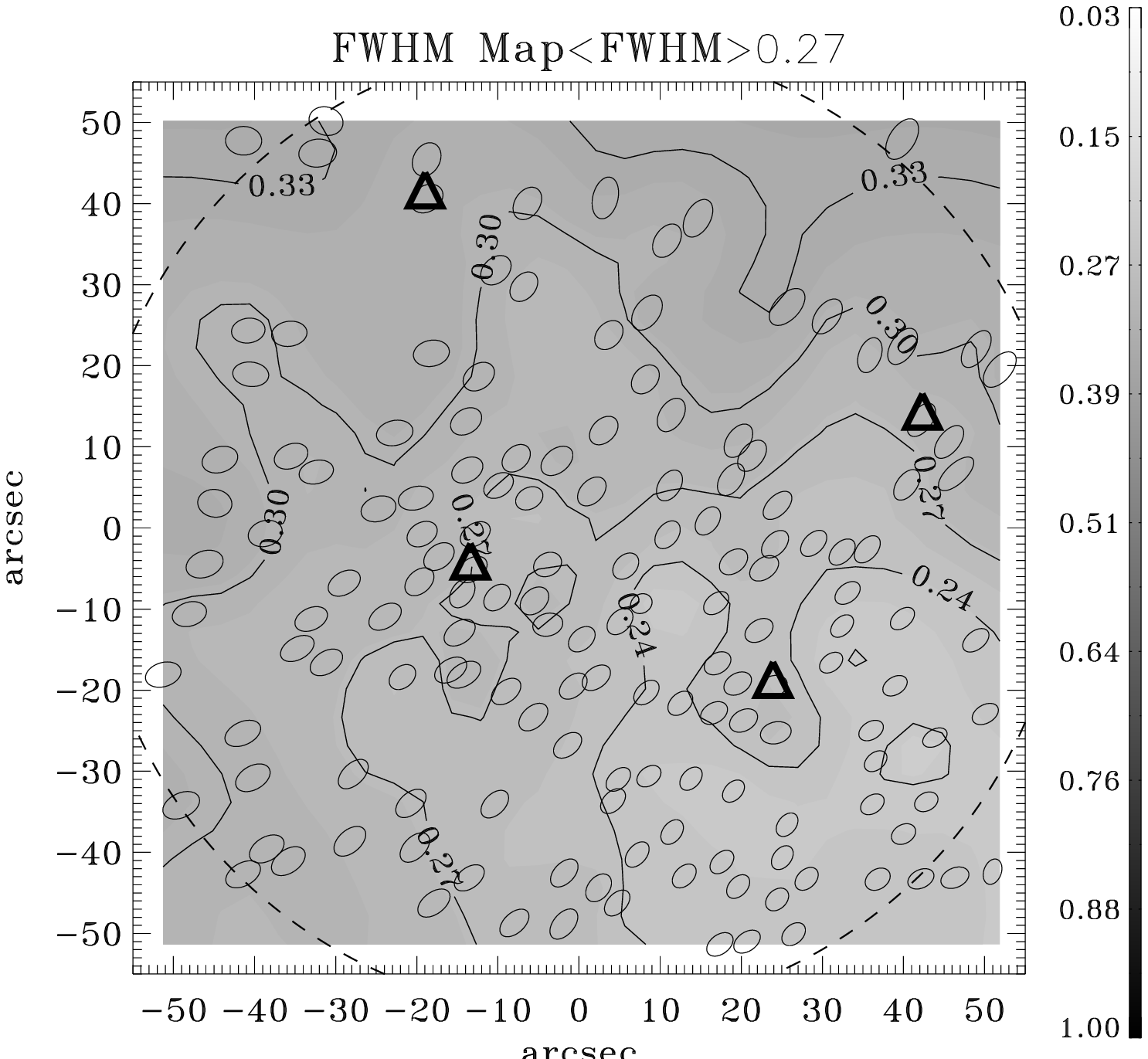}
\includegraphics[width=8cm]{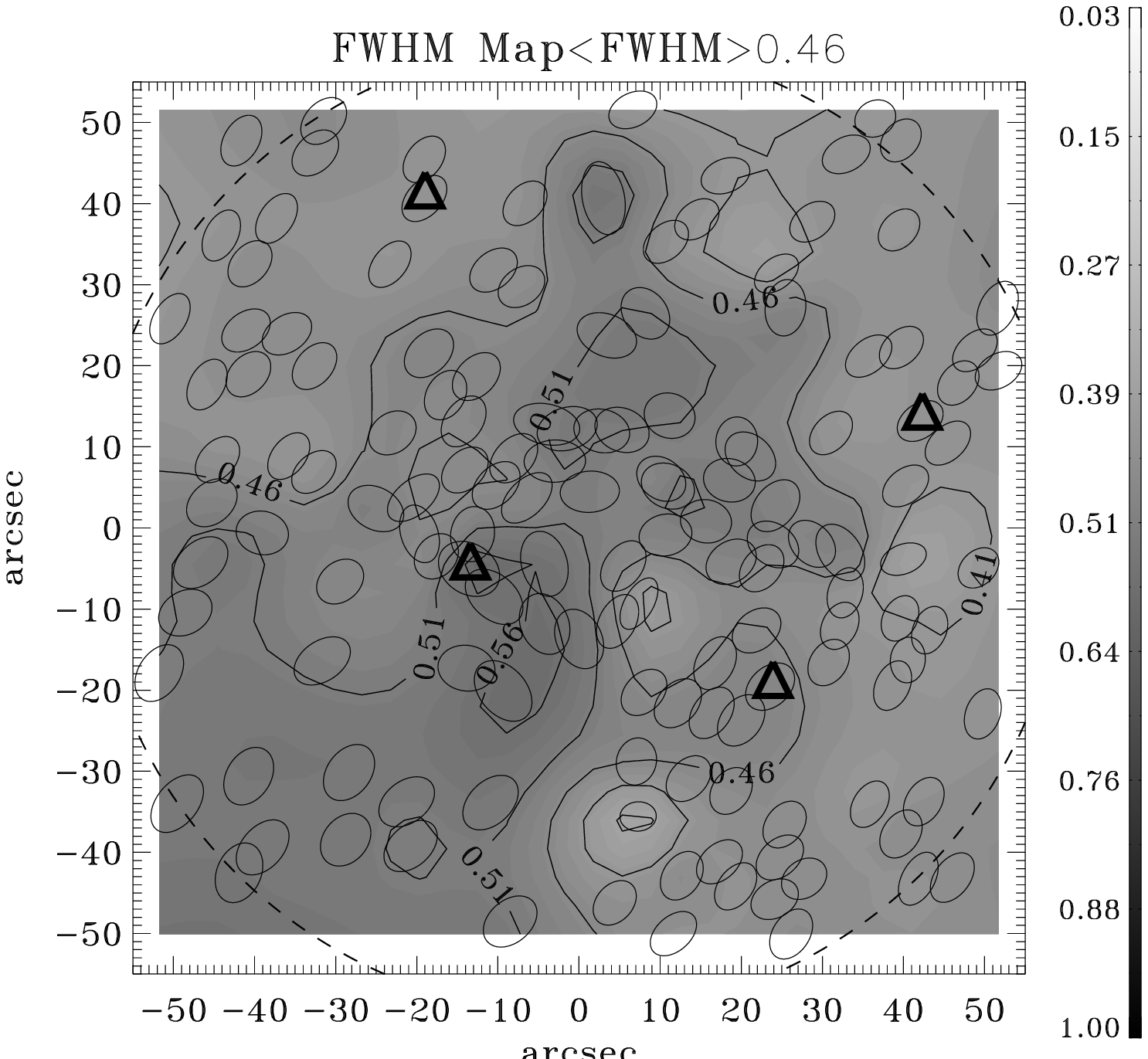}}\caption{\footnotesize
FWHM in Closed (left) and Open (right) loop. The small ellipses represents the positions, orientations and sizes (arbitrarily enlarged for display purposes) of the stars identified and used to compute FWHM and SR values. Triangles indicate the positions of the 4 reference stars.}\label{fig:fwhm}
\end{figure}

\begin{table}[h]
\caption{The following table summarizes the results obtained in GLAO on the Globular Cluster 47Tuc. Full width at half Maximum are measured by fitting Moffat Function (in ${{\rm Br}}\gamma$).  The Ensquared Energy in 0.1\arcsec (EE$_{0.1}$) (${{\rm Br}}\gamma$) is listed showing a gain $\sim \times 2$ with respect open loop case. SR$_{{\rm Br}}\gamma$ and $\sigma_{{\rm{V,DIMM}}}$ are respectively the measured Strehl Ratio and the seeing FWHM measured by the DIMM during the exposure. A mosaic of 5 30seconds exposure frames covers the 2$\times$2 arcmin field. The error has been computed as the standard deviation of the data, while FWHM, EE and SR values are the medians of the measured data. For the presented case 30 over 60 modes have been used in closed loop.}
\label{tab:GLAO0}
\begin{center}
\begin{tabular}{|l|r|r|r|r|r|r|r|} 
\hline
\rule[-1ex]{0pt}{3.5ex}                           & M$_{V}$       & FWHM [\arcsec] ${{\rm Br}}\gamma$      & EE$_{0.1} [\%] $ ${{\rm Br}}\gamma$  &SR$_{{\rm Br}}\gamma$ [\%]& $\sigma_{{\rm{V,DIMM}}}$ [\arcsec] @V\\
\hline
\rule[-1ex]{0pt}{3.5ex} Open Loop \bytwo &  -                     & 0.64$\pm$0.06  &4.1$\pm$1.6  &   1.7$\pm$0.4    & 1.71\arcsec\\
\hline
\rule[-1ex]{0pt}{3.5ex} GLAO Loop \bytwo & 11.9, 11.9, 12.4, 12.5 & 0.38$\pm$0.04  & 8.8$\pm$3.2 &   2.5$\pm$1.0    & 1.54\arcsec\\
\hline
\rule[-1ex]{0pt}{3.5ex} GLAO Loop \byone & 11.9, 11.9, 12.4, 12.5 & 0.21$\pm$0.04  & 12.2$\pm$5.4&   6.1$\pm$3.1    & 1.42\arcsec\\
\hline
\end{tabular}
\end{center}
\end{table}

\section{Multiconjugate Adaptive Optics}
During the nights following the 21$^{\rm st}$ seeing was really poor and no further optimization was possible during the technical nights. During the Guaranteed Time Observation (GTO) nights we had only one chance to test in reasonable conditions the performance of the LO system even if only in faint end observing the globular cluster NGC6388. The median seeing experienced during the whole 9 nights run has been 1.25\arcsec (measured by DIMM seeing monitor and registered on the header of each fits files).
Anyway at the end of the first technical night (the 21$^{\rm st}$) we succeeded to close the MCAO loop even if on a reference stars constellation of only three bright elements (magnitudes 11.059, 11.157 and 12.072).
\begin{table}[h]
\caption{The following table summarizes the results obtained in GLAO and MCAO mode on the first bright MCAO case. Full width at half Maximum, FWHM, are measured by fitting Moffat Function on the ${{\rm Br}}\gamma$ images.  The Ensquared Energy in 0.1\arcsec (EE$_{0.1}$) is listed showing a gain $\sim \times 3$ and $\sim \times 4.5$ with respect open loop case respectively for the GLAO and MCAO. SR$_{{\rm Br}}\gamma$ and $\sigma_{{\rm{V,DIMM}}}$ are respectively the measured Strehl Ratio and the seeing FWHM measured by the DIMM during the exposure (in V-band). The field is a mosaic of 5 exposures (30~seconds each) in order to cover the \bytwo.}
\label{tab:MCAO0}
\begin{center}
\begin{tabular}{|l|r|r|r|r|r|r|r|} 
\hline
\rule[-1ex]{0pt}{3.5ex}                           & M$_{V}$              & FWHM [\arcsec]   ${{\rm Br}}\gamma$       & EE$_{0.1} [\%]$ ${{\rm Br}}\gamma$ &SR$_{{\rm Br}}\gamma$ [\%] & $\sigma_{{\rm{V,DIMM}}}$ [\arcsec] @V\\
\hline
\rule[-1ex]{0pt}{3.5ex} Open Loop               &  -                     & 0.45           &  4.7         &    1.8             &1.48\arcsec\\
\hline
\rule[-1ex]{0pt}{3.5ex} GLAO Closed Loop & 11.059, 11.157, 12.072 & 0.17$\pm$0.02  & 14.9$\pm$2.1 &   9.2$\pm$2.6     & 1.46\arcsec\\
\hline
\rule[-1ex]{0pt}{3.5ex} MCAO Closed Loop & 11.059, 11.157, 12.072 & 0.12$\pm$0.04  & 23.3$\pm$3.9 &  17.3$\pm$9.1     & 1.39\arcsec\\
\hline
\end{tabular}
\end{center}
\end{table}
Regarding globular cluster NGC6388 two different observations have been done: one on the globular cluster center the 26$^{\rm th}$ September (17~36~17.86, -44~44~05.60) and the others on a external Field of the same cluster the 27$^{\rm th}$, 28$^{\rm th}$ and 29$^{\rm th}$ September (RA=17~36~22.86, DEC=-44~45~35.53). Unfortunately only the night of the 27$^{\rm th}$ had acceptable seeing conditions: the results of this night will be discussed in the following while for the others we just say that we obtained only a FWHM improvement and Ensquared Energy concentration gain with respect to open loop seeing PSF, however far in absolute terms from the results obtained in the best night.
\begin{table}[h]
\caption{The following table summarizes the results obtained in MCAO mode on an external region of the Globular Cluster NGC6388. FWHM are measured by fitting Moffat Function.  SR$_{{\rm K'}}$ and $\sigma_{{\rm{V,DIMM}}}$ are respectively the measured Strehl Ratio in SR ${{\rm K'}}$ and the seeing FWHM measured by the DIMM during the exposure in V filter. In this case the M$_{V}$ data are Hubble Space Telescope F606W-filter photometry data. The measured quantities have been estimated on the composite image made by summing the whole 33 K'-frames set, for an overall exposure time of 490 seconds.}
\label{tab:MCAO1}
\begin{center}
\begin{tabular}{|l|r|r|r|r|r|r|r|} 
\hline
\rule[-1ex]{0pt}{3.5ex}                           & M$_{V}$              & FWHM [\arcsec] @K' & EE$_{0.1} [\%]$ @K' &SR$_{{\rm K'}}$ [\%] & $\sigma_{{\rm{V,DIMM}}}$ [\arcsec] @V\\
\hline
\rule[-1ex]{0pt}{3.5ex} Faint MCAO Closed Loop & 15, 15, 15.6, 15.7, 16.2 & 0.15$\pm$0.01  & 19.9$\pm$4.9 &   10.8$\pm$2.1     &  0.55 \\
\hline
\end{tabular}
\end{center}
\end{table}

\begin{figure}
\centerline{\includegraphics[width=8cm]{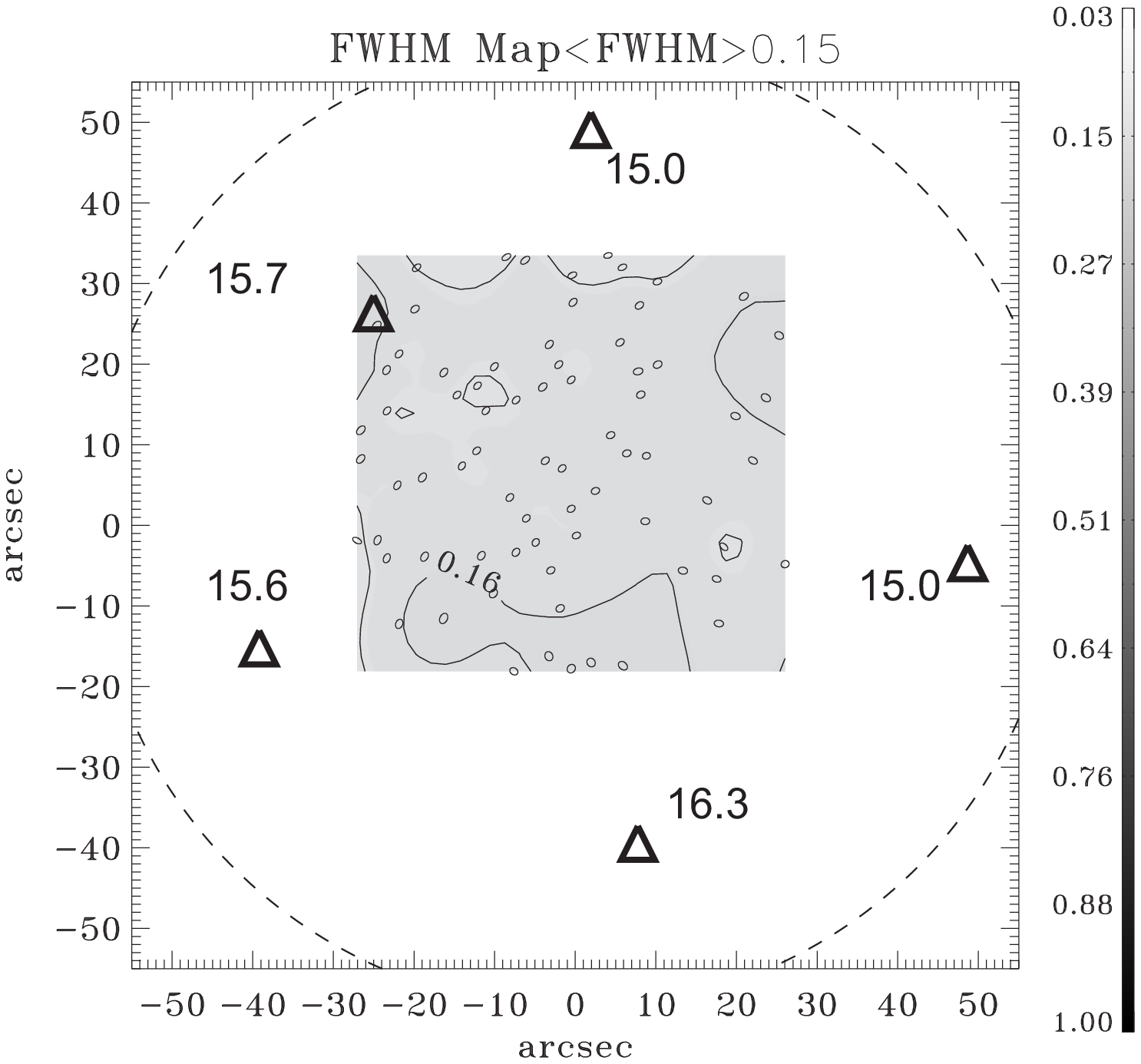}
\includegraphics[width=8cm]{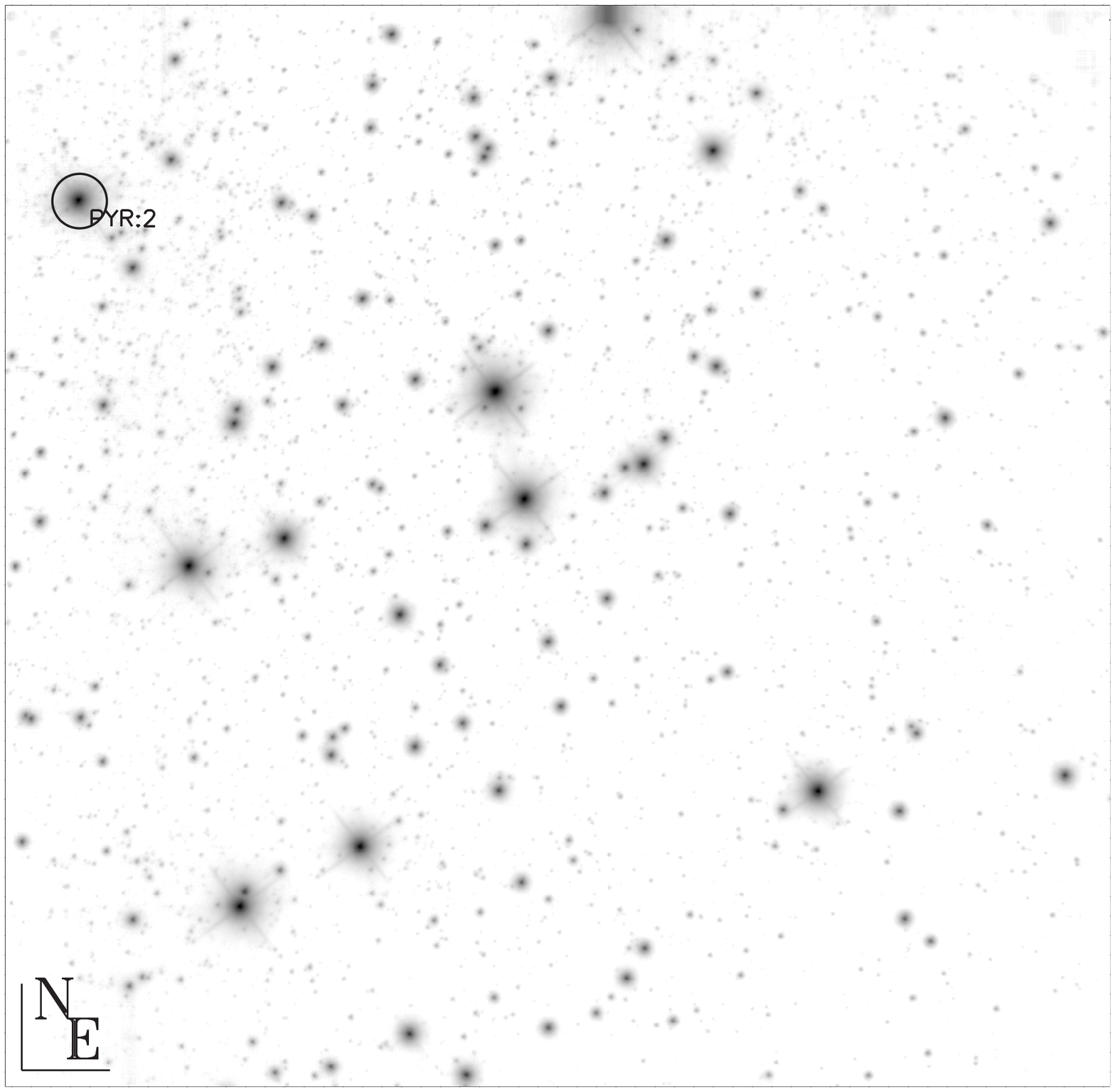}}\caption{\footnotesize
On the left picture the positions of the reference stars (triangles) with respect to the observed FoV are superimposed to the measured FWHM map. Close to the reference stars is overwritten the corresponding HST-F606W filter magnitude. On this region the FWHM resulted to be close to diffraction limited nevertheless the faint magnitudes of the reference stars, which integrated magnitude is 13.67. The ellipses represents the size (enlarged) and orientation of the PSF such as been fitted using Moffat Functions. On the right an image of the field with superimposed a circle indicating one of the 5 reference stars.}\label{fig:6388a}
\end{figure}

\begin{figure}
\centerline{\includegraphics[width=8.2cm]{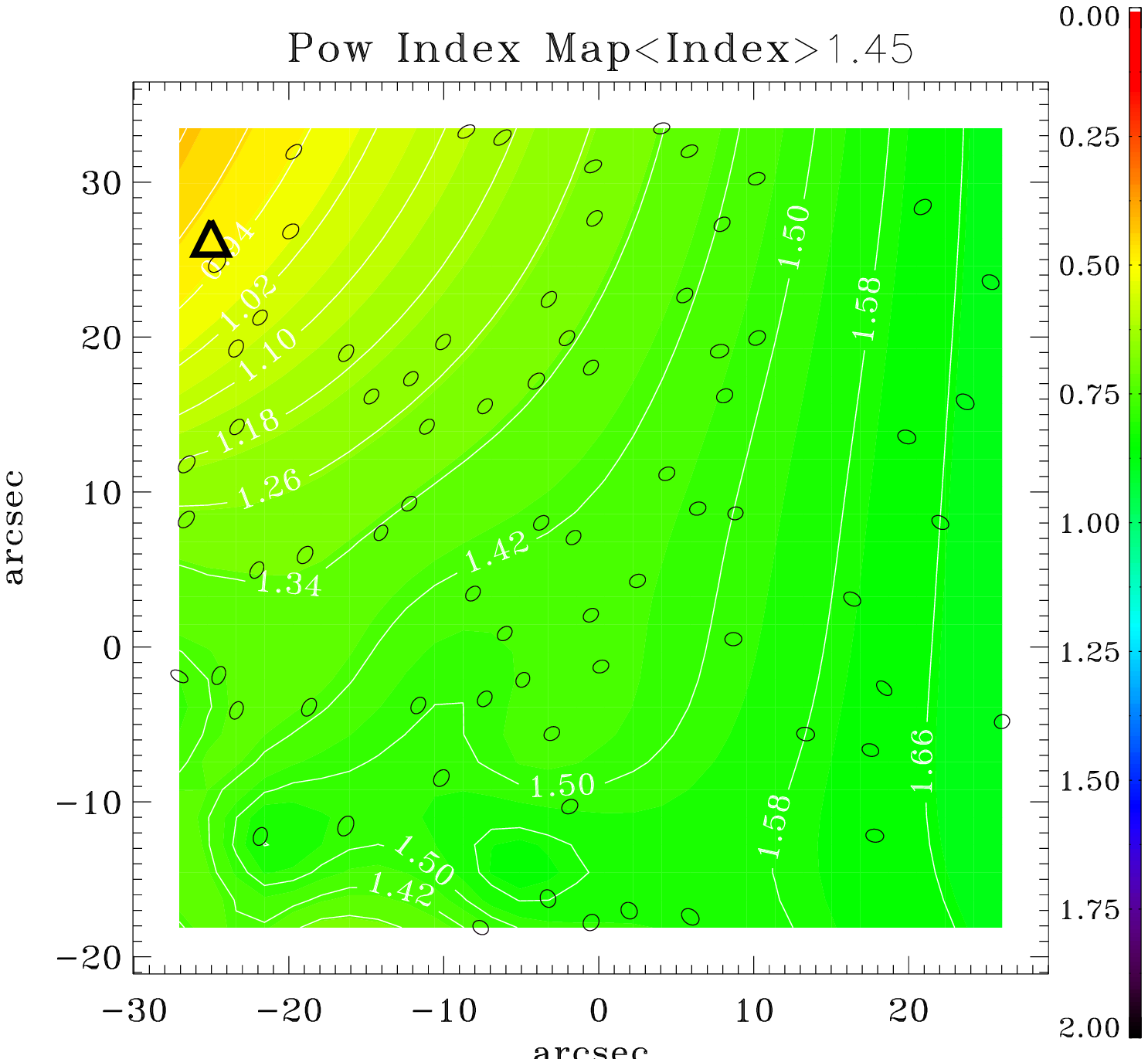}
\includegraphics[width=8.2cm]{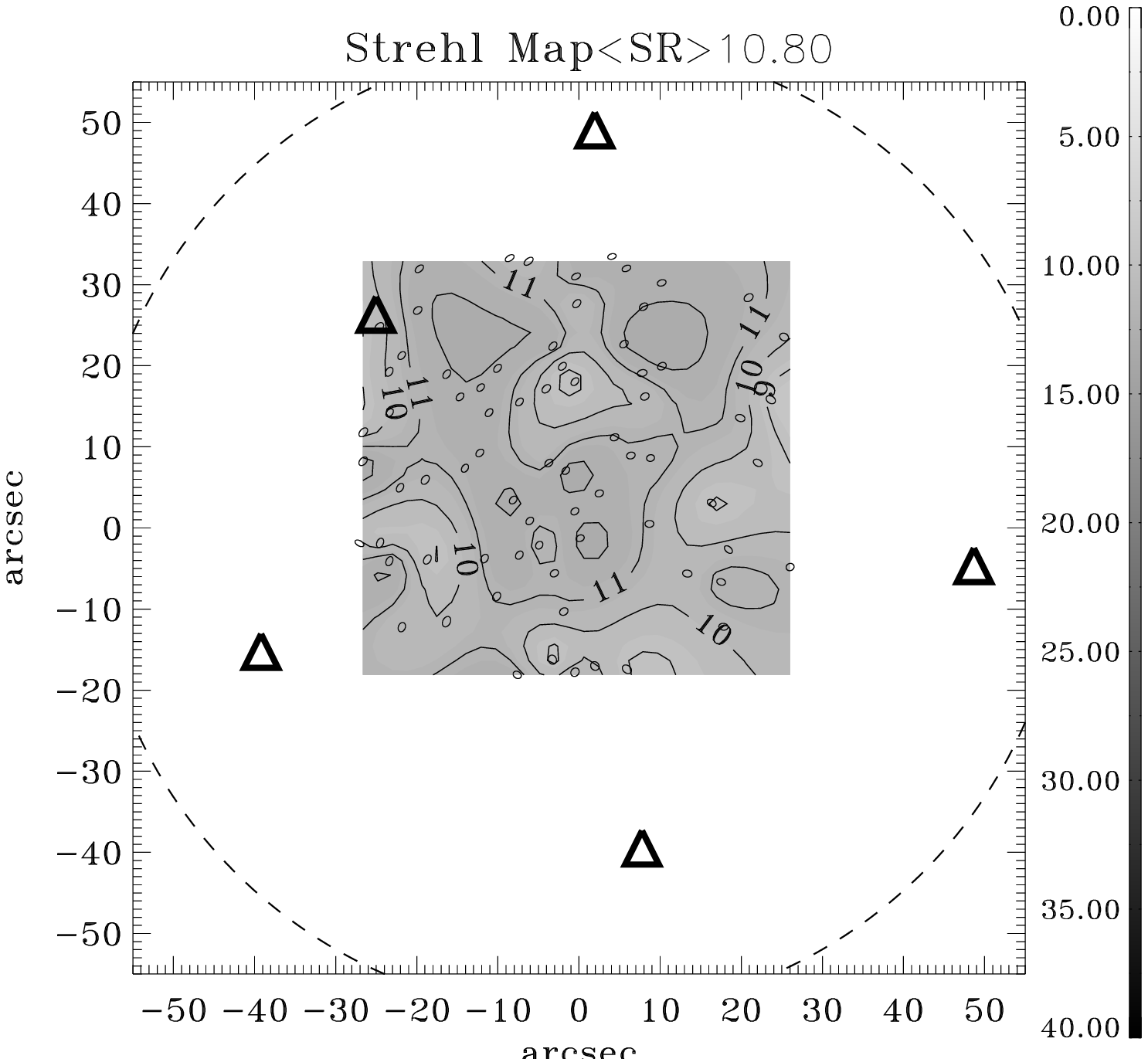}}\caption{\footnotesize
    The left picture shows the Moffat Power law index distribution. Close to the reference stars this quantity has its minimum. The ellipses represents the size and orientation of the PSFs such as been fitted using Moffat Function. On the right the Strehl Ratio distribution. Both are measured on K' filter images.}\label{fig:6388b}
\end{figure}

\section{A few notes about PSF}
\begin{figure}
\centerline{\includegraphics[width=8cm]{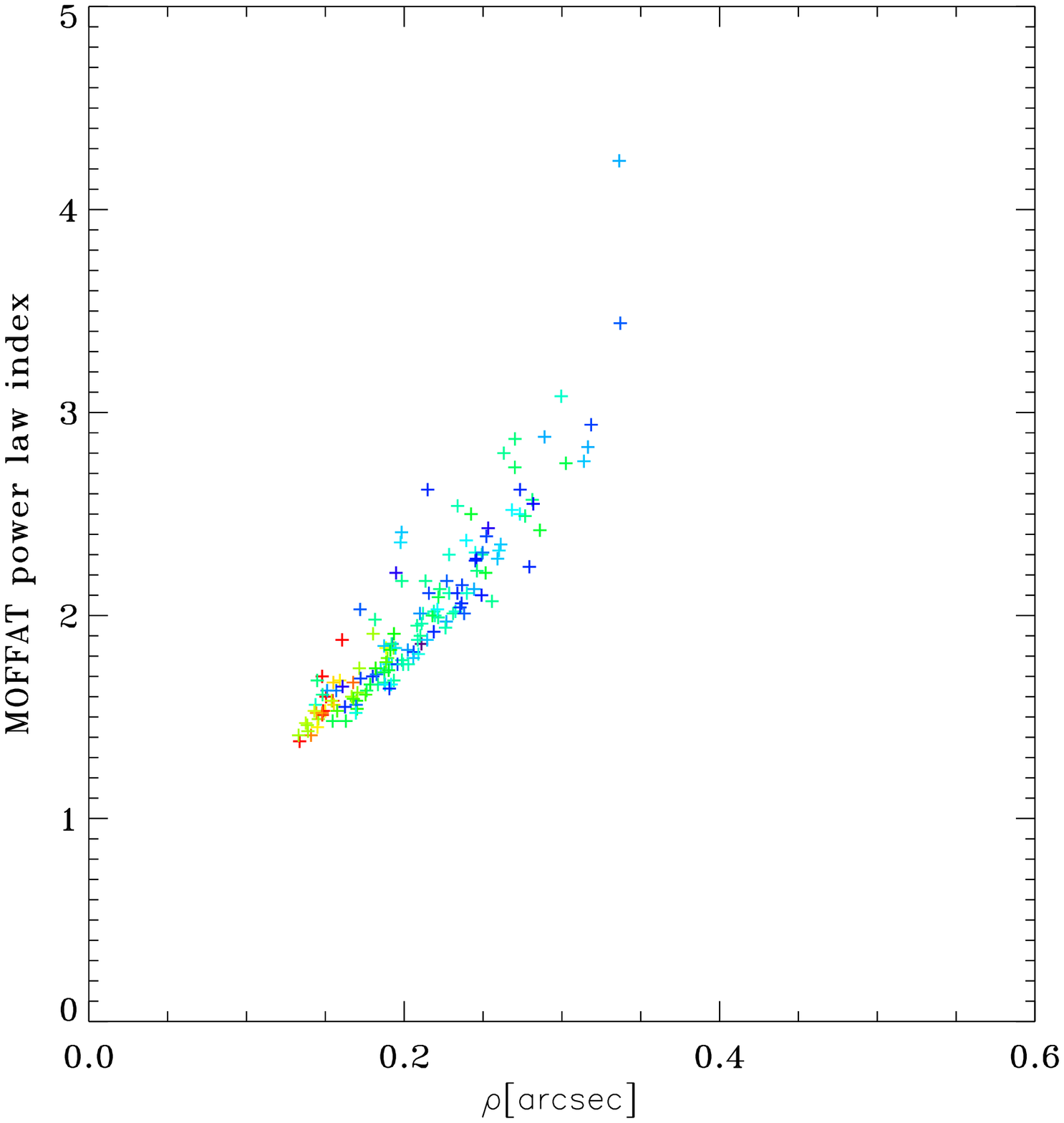}
\includegraphics[width=8cm]{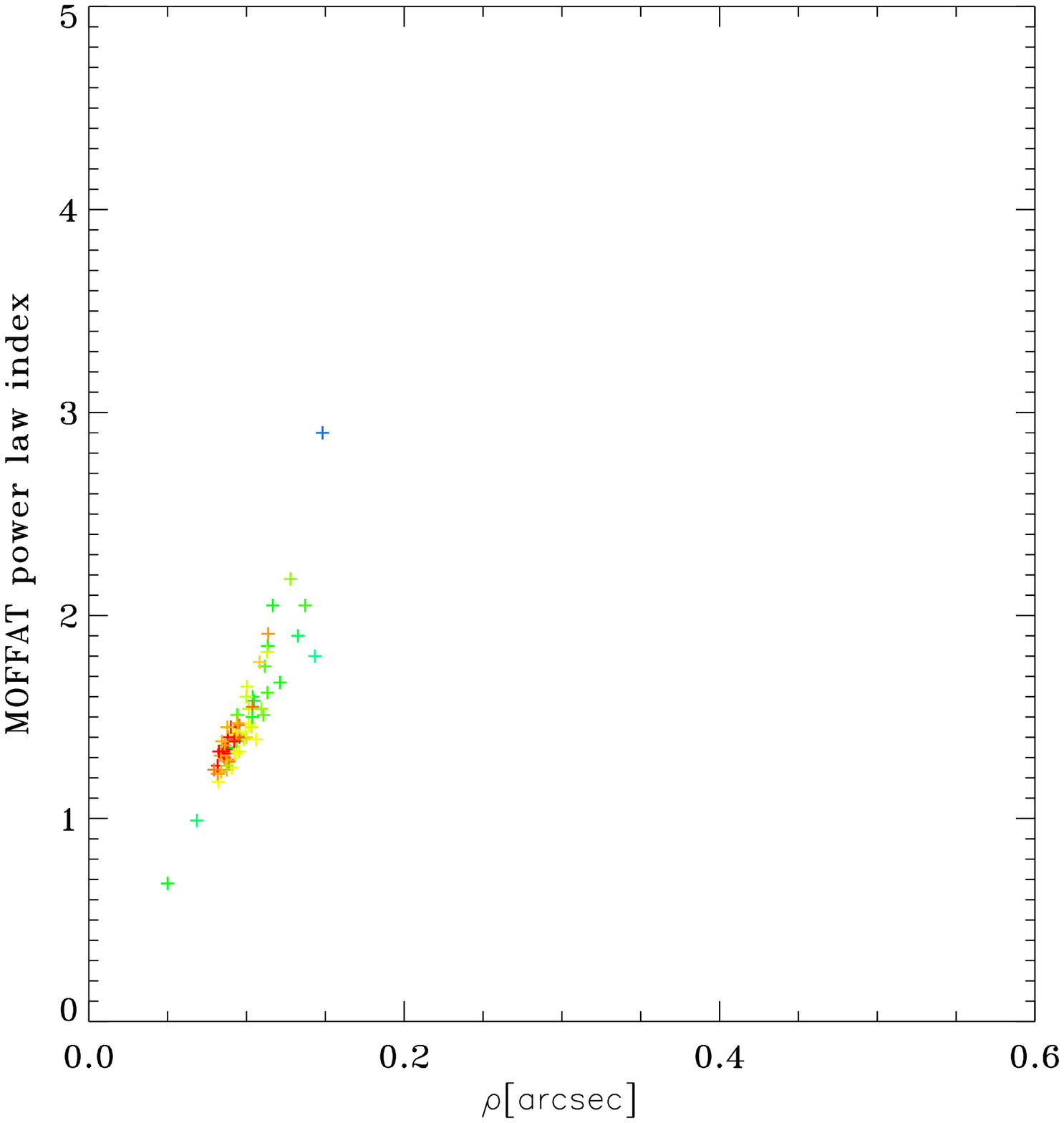}}\caption{\footnotesize
In the plot above we collected the data relative to PSF Moffat fit. On the left the case discussed about GLAO correction on 47Tuc on the right the MCAO NGC6388 case. The linear relation between the two parameters is clearly visible. SR in these plots goes from the top right to the bottom left following the linear relation.}\label{fig:rhobeta}
\end{figure}

Since we had available the MAD MCAO data we studied the problem of the PSFs uniformity over the FoV and its characteristics, especially important in the case of crowded field such as the globular clusters. In crowded field it's preferable to use PSF fitting instead then aperture photometry in order to measure the flux relative to each source. But in the case of the fitting it's important to use a proper fitting function. The widely used DAOPHOT\cite{daophot,stetson90} fits PSF using Penny or Moffat functions (or other variations and combinations of the two). Tests we performed that we do not report here show that in the GLAO and MCAO case (and more generally AO) the PSF are composed by two main structures: the core (the coherent light close to the diffraction limited peak) and the halo (the uncorrected turbulence). Over these two components the diffraction pattern becomes important at medium-high Strehl Ratio values (more than 20\%-30\%).
Moreover in many cases static aberrations such as non common path aberrations on the side of camera optical train introduce a fixed (and by chance small) pattern, which varies over the field. In practice in the case of low SR PSF fitting using common Penny or Moffat function works pretty well, but going to medium SR regime, between 10\% and 30\%, the PSF is composed by to components which can be fitted using Penny function or composition of two different Moffats. On higher SR the diffraction pattern and static aberrations should be taken into account.
As example of the intriguing studies possible we show in the Figure~\ref{fig:rhobeta} the relation between the parameters describing the Moffat\cite{moffat} function:
\begin{equation}
f(x,y) = c + {\rm I}_{\rm 0} \left[ \left( \frac{x-x_{\rm 0}}{\rho_{\rm x}} \right)^2 + \left(\frac{y-y_{\rm 0}}{\rho_{\rm y}} \right)^2 + 1\right]^{-\beta}
\label{eq:moffat}
\end{equation}
where I$_{\rm 0}$ is the intensity, x$_{\rm 0}$ and y$_{\rm 0}$ are the xy coordinates of the center, $\beta$ the power law index parameter and $\rho$ another shape parameter.

From the data analysis clearly is visible a common behavior:
in a $\rho$ versus $\beta$ plot (see Figure~\ref{fig:rhobeta}) data collected on certain atmospheric seeing tend to be distributed
according to a linear relation more and more steep smaller the open loop seeing. This relation is due to the adaptive optics: the AO corrects up to a spatial frequency, which depends on the spatial resolution of the deformable mirror and of the WFS.
High spatial frequencies, which are responsible of PSF wings, are only partially smoothed: in energy terms AO moves photons from PSF wings into the central core, flattening wings and stretching the peak. Weaker is the initial turbulence, smaller is open loop seeing disk size, fastest is the relation $\beta$ vs $\rho$ being smaller the energy fraction on the PSF halo. The atmospheric turbulence power spectrum fixes the energy distribution, AO cuts low order-spatial frequencies: the halo intensity  smaller the Strehl Ratio higher.

\section{Conclusions}
The Layer Oriented concept has been demonstrated working also on sky and not only in laboratory. The actual performance did not been addressed because of the poor seeing condition during the whole observations run. Nevertheless the possibility to use very faint (in AO standards) stars for wavefront sensing has been demonstrated. Moreover also in bad seeing condition the system well performed even if not with resolution close to diffraction limit but working in a seeing reduction regime with a gain in Ensquared Energy of the order of a factor 2$\times$to 3$\times$ in 0.1\arcsec. This multi-pyramids scheme has the possibility to reach non negligible sky coverage at the galactic poles using only natural guide stars if several corrections with respect to the MAD design will be done: such as WFS CCD sensible to R- or IR- bands instead of B/V, modulation of the pyramids and proper obscuration of the sky background. The firsts scientific papers obtained during the described observation run have been published\cite{Gullieuszik08,mignani08} and soon will be followed by others. Stay tuned...

\acknowledgments     
We would acknowledge all the people which have let the Layer Oriented for MAD possible and with their support helped us to make the Layer Oriented WFS finally (and properly) working: P. Amico, R. Gilmozzi, W. G\"assler, M. Le Louarn, P. Salinari, C. Dupuy, A. Moretti, M. Gullieuszik, R. Falomo and M. Carbillet among the others.


\bibliography{spie}   
\bibliographystyle{spiebib}   

\end{document}